\documentclass[preprint, aip,jmp]{revtex4-1}

\usepackage{amssymb}
\usepackage{amsmath}
\usepackage{amscd}
\usepackage{natbib}
\usepackage{verbatim}
\def\d{\delta}

\def\be{\begin{equation}}
\def\ee{\end{equation}}
\def\arr{\begin{array}{rll}}
	\def\ea{\end{array}}
\def\bea{\begin{eqnarray}}
\def\eea{\end{eqnarray}}

\def\N2{$N{=}2$}

\def\>{\rangle}
\def\<{\langle}
\def\+{\dagger}
\def\={\ =\ }

\begin{document}
	\renewcommand{\thefootnote}{\arabic{footnote}}
	
	\title{Higher-derivative generalization of conformal mechanics}
	
	\author{Oleg Baranovsky}
	
	\affiliation{Laboratory of Mathematical Physics, Tomsk Polytechnic University, 634050 Tomsk, Lenin Ave. 30, Russian Federation}
	
	\email{oleg.bbaranovskiy@gmail.com}
	
	\begin{abstract}
		Higher-derivative analogues of multidimensional conformal particle and many-body conformal mechanics are constructed. Their Newton-Hooke counterparts are derived by applying appropriate coordinate transformations.
	\end{abstract}
	
	
	
	\maketitle
	
	{\bf 1. Introduction}
	\vskip 0.5cm
	Conformal algebra in one dimension $so(1,2)$ involves three generators: the generator of time
	translations $H$, the generator of dilatations $D$, and the generator of special conformal transformations $K$.
	There are many mechanical systems which exhibit such a symmetry. These include the model of free particle \cite{Niederer_1}, conformal particle \cite{Alfaro},
	harmonic oscillator \cite{Niederer}, the system of identical particles which interact with each other by conformal potential (e.g. the Calogero
	model \cite{Calogero}) and others.
	
	There are several reasons why conformal invariant mechanical systems have attracted interest over the last
	four decades. First, some of these models are (super)integrable. Second, there are conformal invariant systems
	which are of physical interest. In particular, some of these models appear in the context of condensed matter physics \cite{Johnson},
	in physics of black holes \cite{Gibbons}. Third, recent proposals on the nonrelativistic version of the AdS/CFT-correspondence stimulate
	investigations of nonrelativistic conformal algebras and their dynamical realizations.
	
	There are two families of nonrelativistic algebras which contain $so(1,2)$ as a subalgebra. The first family involves
	conformal extensions of the Galilei algebra which are parameterized by a positive integer or half-integer parameter $l$
	\cite{Horvathy,Henkel,Negro_1,Negro_2,Lukierski}. This is the reason why the members of this family are called $l$-conformal Galilei algebras.
	
	The $l$-conformal extensions of the Newton-Hooke (NH) algebra form the second family of nonrelativistic conformal algebras \cite{Negro_1,Galajinsky_3}. The $l$-conformal NH algebra can be viewed as an analogue of the $l$-conformal Galilei algebra in the
	presence of universal cosmological repulsion or attraction.
	
	As is well known, the $l$-conformal extensions of the Galilei algebra and the NH algebra are isomorphic. Indeed, the structure relations of
	the $l$-conformal NH algebra can be obtained by a change of the basis $H\rightarrow H\pm\frac{1}{R^2}K$ in the $l$-conformal Galilei algebra, where
	$\Lambda=\pm\frac{1}{R^2}$ is the nonrelativistic cosmological constant. But when dynamical realizations are considered, this change of
	the basis leads to a change of the Hamiltonian and consequently alters the dynamics. For example, the Schr\"{o}dinger group ($l=1/2$-conformal
	Galilei group) is the maximal kinematical group of the free particle \cite{Niederer_1}. At the same time, the NH counterpart of this model is the harmonic oscillator \cite{Niederer}.
	
	The free higher-derivative particle \cite{DH,Gomis} and the Pais-Uhlenbeck (PU) oscillator \cite{Pais} can be viewed as higher-derivative analogues of the free particle and
	the harmonic oscillator, respectively. Recently, symmetries of these higher-derivative models have been extensively studied \cite{DH,Gomis,Galajinsky_1,AG,PU,Andrzejewski,Andrzejewski_1,Masterov_2,AB,Andr}. 
	In particular it has been shown that the $l$-conformal Galilei group is the maximal symmetry group
	of the free $(2l+1)$-order particle \cite{Gomis}. Similarly, the PU oscillator accommodates the $l$-conformal NH symmetry for a special choice of  its oscillation frequencies \cite{Galajinsky_1,PU,Masterov_2}.
	
	At the same time, higher-derivative analogues of other nonrelativistic conformal invariant mechanical systems are unknown. The purpose of the present work is to
	construct higher-derivative analogues of the conformal particle and a
	system of identical particles which interact with each other via a conformal-invariant potential.
	
	The paper is organized as follows. In Sect. 2, we review the free higher-derivative particle and the PU oscillator
	which exhibits the $l$-conformal NH symmetry. In Section 3 and Section 4, we construct higher-derivative analogues of the conformal particle and
	many-body conformal mechanics, respectively. We summarize our results and discuss further possible developments in the conclusion (Section 5).
	Throughout the work summation over repeated spatial indices is understood. A superscript in braces as well as the number of dots over spatial
	coordinates designate the number of derivatives with respect to time.

	\vskip 0.5cm
	\noindent
	{\bf 2. Free higher-derivative particle and the Pais-Uhlenbeck oscillator}
	\vskip 0.5cm
	Let us review certain useful facts about the free higher-derivative particle and the PU oscillator which enjoys the $l$-conformal NH symmetry.
	The action functional of the former model reads \cite{Gomis}
	\bea\label{FHP}
	S=\frac{1}{2}\int dt\,\lambda_{ij}x_i x_j^{(2l+1)},
	\eea
	where dimensionless parameter $l$ can take positive integer or half-integer values and
	\bea
	\lambda_{ij}=\left\{
	\begin{aligned}
		& \delta_{ij}, && \quad i,j=1,2,..,d, \quad \mbox{for half-integer $l$;}
		\\[2pt]
		& \epsilon_{ij}=-\epsilon_{ji}, && \quad i,j=1,2, \qquad \quad \qquad \mbox{for integer $l$},
	\end{aligned}
	\right.
	\nonumber
	\eea
	with $\epsilon_{12}=1$. The action functional (\ref{FHP}) is invariant under the transformations \cite{Gomis}
	\bea
	&&
	t'=t+a+bt^2+ct,\quad {x'}_i(t')=x_i(t)+\sum_{k=0}^{2l}a_i^{(k)}t^k+lcx_i(t)+2lbtx_i(t)-\omega_{ij}x_j(t),
	\nonumber
	\eea
	where $a$, $b$, $c$, $a_i^{(k)}$, and $\omega_{ij}$ are infinitesimal parameters. Generators of these transformations
	form the $l$-conformal Galilei algebra \cite{Henkel,Negro_1,Negro_2}.
	
	The NH counterpart of the model (\ref{FHP}) can be constructed by applying the Niederer transformation \cite{Niederer}
	\bea\label{Nied}
	t=R \tan{(\tau/R)},\quad x_i(t)=\frac{\chi_i(\tau)}{\cos^{2l}{(\tau/R)}},
	\eea
	where $R$ is a constant which has the dimension of time. For half-integer $l$, the implementation of this transformation to (\ref{FHP}) results in the action
	functional \cite{Galajinsky_1,PU}
	\bea\label{PUev}
	S=\frac{1}{2}\int d\tau\, L_{PU}=\frac{1}{2}\int d\tau\,\chi_i\prod_{k=0}^{l-1/2}\left(\frac{d^2}{d\tau^2}+\frac{(2k+1)^2}{R^2}\right)\chi_i,
	\eea
	while for integer $l$ one has \cite{Masterov_2}
	\bea\label{PUod}
	S=\frac{1}{2}\int d\tau\, L_{PU}=\frac{1}{2}\int d\tau\,\epsilon_{ij}\,\chi_i\prod_{k=1}^{l}\left(\frac{d^2}{d\tau^2}+\frac{(2k)^2}{R^2}\right)\dot{\chi}_j.
	\eea
	The actions functionals (\ref{PUev}) and (\ref{PUod}) describe the $(2l+1)$-order PU oscillator for a particular choice of its
	oscillation frequencies. This model is invariant under the transformations \text{   }       \cite{Galajinsky_1,PU,Masterov_2}
	\bea\label{tr4}
	\begin{aligned}
		&
		\tau'=\tau+a+\frac{cR}{2}\sin{\frac{2\tau}{R}}+bR^2\sin^2{\frac{\tau}{R}},
		\\[2pt]
		&
		{\chi'}_i(\tau')=\chi_i(\tau)+lc\cos{\frac{2\tau}{R}}\chi_i(\tau)+lbR\sin{\frac{2\tau}{R}}\chi_i(\tau)+
		\sum_{k=0}^{2l}R^k\sin^{k}{\frac{\tau}{R}}\cos^{2l-k}{\frac{\tau}{R}}a_i^{(k)}-\omega_{ij}\chi_j(\tau),
	\end{aligned}
	\eea
	whose generators form the $l$-conformal Newton-Hooke algebra \cite{Negro_1,Negro_2,Galajinsky_3}. 
	
	The NH counterparts presented in this paper all correspond to the case of negative cosmological constant. The case of positive cosmological constant can be straightforwardly reproduced by a formal change $R\rightarrow iR$. 
	
	\vskip 1cm
	\noindent
	{\bf 3. Higher-derivative analogue of conformal particle}
	\vskip 0.5cm
	
	Let us consider a multidimensional conformal particle whose action functional has the form \cite{Alfaro}
	\bea\label{CP}
	S=\frac{1}{2}\int dt \left(\dot{x}_i\dot{x}_i-\frac{g}{x_i x_i}\right).
	\eea
	This action is invariant under transformations
	\bea
	t'=t+a+ct+bt^2,\quad {x'}_i(t')=x_i(t)+\frac{c}{2}x_i(t)+bt x_i(t)-\omega_{ij}x_j(t),\quad \omega_{ij}=-\omega_{ji},
	\nonumber
	\eea
	whose generators form $so(1,2)\oplus so(d)$ subalgebra of the $l=1/2$-conformal Galilei algebra. For the model of a free higher-derivative particle
	(\ref{FHP}), the same subalgebra is realized by the generators which produce the following transformations
	\bea\label{tr1}
	t'=t+a+ct+bt^2,\quad {x'}_i(t')=x_i(t)+lc x_i(t)+2lbt x_i(t)-\omega_{ij}x_j(t),\quad \omega_{ij}=-\omega_{ji}.
	\eea
	
	Taking into account (\ref{FHP}), let us consider the action functional of
	the form
	\bea
	S=\frac{1}{2}\int dt \left(\lambda_{ij}x_i x_j^{(2l+1)}-V(x)\right)
	\nonumber
	\eea
	with an arbitrary function $V=V(x)$. To obtain higher-derivative generalization of the model (\ref{CP}), let us require this action
	functional to be invariant under the transformations (\ref{tr1}). This restriction is satisfied
	when the function $V=V(x)$ has the form
	\bea
	V(x)=\frac{g}{(x_i x_i)^{1/2l}}.
	\nonumber
	\eea
	So, the model
	\bea\label{CHP}
	S=\frac{1}{2}\int dt \left(\lambda_{ij}x_i x_j^{(2l+1)}-\frac{g}{(x_i x_i)^{1/2l}}\right)
	\eea
	can be viewed as a higher-derivative generalization of conformal particle (\ref{CP}). The dynamics of this system is governed by
	the following equations of motion
	\bea\label{equ}
	\lambda_{ij}x_j^{(2l+1)}=-\frac{g}{2l}\frac{x_i}{(x_j x_j)^{(2l+1)/2l}}.
	\eea
	
	It is interesting to note that a one-dimensional analogue of this dynamical equation can be obtained via the method
	of nonlinear realization \cite{Coleman_1,Coleman_2,Ivanov}. Indeed, let us consider the exponential parametrization of the group $SO(1,2)$ \cite{Ivanov_1}
	\bea
	G=G(t,z,u)=e^{itH}e^{izK}e^{iuD},\quad [H,D]=iH,\; [H,K]=2iD,\; [D,K]=iK.
	\nonumber
	\eea
	Left multiplication by a group element $G(a,b,c)$ produces the following infinitesimal coordinate
	transformations
	\bea\label{tr2}
	\d t=a+bt^2+ct,\quad \d z=b(1-2tz)-cz,\quad \d u=c+2bt,
	\eea
	where $a$, $b$, and $c$ are infinitesimal parameters. Then one constructs the left-invariant Maurer-Cartan one-forms \cite{Ivanov_1}
	\bea
	G^{-1}dG=i(\omega_H H+\omega_D D+\omega_K K),
	\nonumber
	\eea
	where we denoted
	\bea
	\omega_H=e^{-u}dt,\quad \omega_D=du-2zdt,\quad \omega_K=e^u(dz+z^2 dt).
	\nonumber
	\eea
	
	To obtain higher-derivative dynamical realization of $SO(1,2)$ group, firstly let us introduce the new variable
	\bea
	\rho=e^{lu}.
	\nonumber
	\eea
	During next step, we discard the variable $z$ from our consideration with the aid of the constraint
	\bea
	\omega_D=0\;\Rightarrow\; z=\frac{1}{2l}\frac{\dot{\rho}}{\rho}.
	\nonumber
	\eea
	Considering this relation, the transformations (\ref{tr2}) may be rewritten as
	\bea\label{tr3}
	\d t=a+bt^2+ct,\quad \d\rho=lc\rho+2lbt\rho.
	\eea
	
	One may obtain $SO(1,2)$-invariant higher-derivative equations by using both the function
	\bea\label{Omega}
	\Omega=\frac{\omega_K}{\omega_H}=\frac{1}{2l}\rho^{\frac{2}{l}}\left(\frac{\ddot{\rho}}{\rho}-\frac{2l-1}{2l}\frac{\dot{\rho}^2}{\rho^2}\right)
	\nonumber
	\eea
	and the differential operator
	\bea\label{D}
	D=\rho^{\frac{1}{l}}\frac{d}{dt},
	\eea
	which are invariant under the transformations (\ref{tr3}). In particular, one-dimensional analogue of (\ref{equ}) for $l=1,\frac{3}{2},2,\frac{5}{2}$ can be reproduced as follows
	\bea
	\begin{aligned}
		&
		l=1: && D\Omega=-\frac{g}{4}\;\Rightarrow\;\dddot{\rho}=-\frac{g}{2\rho^2};
		\\[2pt]
		&
		l=\frac{3}{2}: && D^2\Omega+3\Omega^2=-\frac{g}{9}\;\Rightarrow\;\rho^{(4)}=-\frac{g}{3\rho^{5/3}};
		\\[2pt]
		&
		l=2: && D^3\Omega+16\Omega D\Omega=-\frac{g}{16}\;\Rightarrow\;\rho^{(5)}=-\frac{g}{4\rho^{3/2}};
		\\[2pt]
		&
		l=\frac{5}{2}: && D^4\Omega+31\Omega D^2\Omega+26(D\Omega)^2+45\Omega^3=-\frac{g}{25}\;\Rightarrow\;\rho^{(6)}=-\frac{g}{5\rho^{7/5}}.
	\end{aligned}
	\nonumber
	\eea
	
	It should be noted that the NH counterpart of the model (\ref{CHP}) can be obtained via a Niederer transformation (\ref{Nied}). The action functional of 
	this counterpart has the form
	\bea\label{CPU}
	S=\frac{1}{2}\int d\tau \left(L_{PU}-\frac{g}{(\chi_i\chi_i)^{1/2l}}\right),
	\eea
	where $L_{PU}$ is defined in (\ref{PUev}) and (\ref{PUod}). This model holds invariant under $SO(1,2)\oplus SO(d)$-subgroup of transformations (\ref{tr4}).
	
	The system (\ref{CPU}) can be also viewed as a higher-derivative analogue of the conformal particle in a harmonic trap
	\bea\label{trap}
	S=\frac{1}{2}\int d\tau \left(\dot{\chi}_i\dot{\chi}_i-\frac{1}{R^2}\chi_i\chi_i-\frac{g}{\chi_i \chi_i}\right).
	\eea
	This action functional can be obtained by applying (\ref{Nied}) to (\ref{CP}).
	
	\vskip 1cm
	\noindent
	{\bf 4. Higher-derivative analogue of many-body conformal mechanics}
	\vskip 0.5cm
	
	A system of $N$ identical particle whose dynamics is governed by the action functional
	\bea
	S=\frac{1}{2}\int dt\left( \sum_{\alpha=1}^{N}\dot{x}_{\alpha,i}\dot{x}_{\alpha,i}-V_{\frac{1}{2}}(x_{1,i},x_{2,i},..,x_{N,i})\right) 
	\nonumber
	\eea
	may exhibit invariance under transformations
	\bea
	t'=t+a+bt^2+ct,\quad {x'}_{\alpha,i}(t')=x_{\alpha,i}(t)+a_i^{(0)}+a_i^{(1)}t+\frac{c}{2}x_{\alpha,i}(t)+btx_{\alpha,i}(t)-\omega_{ij}x_{\alpha,j}(t),
	\nonumber
	\eea
	whose generators form the Schr\"{o}dinger algebra when the function $V_{\frac{1}{2}}=V_{\frac{1}{2}}(x_{1,i},x_{2,i},..,x_{N,i})$ satisfies the following
	equations
	\bea\label{sys1}
	\sum_{\alpha=1}^{N}\frac{\partial V_{\frac{1}{2}}}{\partial x_{\alpha,i}}=0,\quad
	\sum_{\alpha=1}^{N}x_{\alpha,i}\frac{\partial V_{\frac{1}{2}}}{\partial x_{\alpha,i}}+2V_{\frac{1}{2}}=0,\quad
	\sum_{\alpha=1}^{N}\left(x_{\alpha,i}\frac{\partial V_{\frac{1}{2}}}{\partial x_{\alpha,j}}-x_{\alpha,j}\frac{\partial V_{\frac{1}{2}}}{\partial x_{\alpha,i}}\right)=0.
	\eea
	
	To construct the higher-derivative analogue of this model, let us consider the following action functional:
	\bea\label{MB}
	S=\frac{1}{2}\int dt \left(\sum_{\alpha=1}^{N}\lambda_{ij}x_{\alpha,i} x_{\alpha,j}^{(2l+1)}-V_{l}(x_{1,i},x_{2,i},..,x_{N,i})\right).
	\eea
	This action is invariant under the transformations
	\bea
	&&
	t'=a+bt^2+ct,\quad {x'}_{\alpha,i}(t')=x_{\alpha,i}(t)+\sum_{k=0}^{2l}a_i^{(k)}t^k+lcx_{\alpha,i}(t)+2lbtx_{\alpha,i}(t)-\omega_{ij}x_{\alpha,j}(t).
	\nonumber
	\eea
	when the function $V_{l}=V_{l}(x_{1,i},x_{2,i},..,x_{N,i})$ obeys the following equations
	\bea\label{sys2}
	\sum_{\alpha=1}^{N}\frac{\partial V_{l}}{\partial x_{\alpha,i}}=0,\quad
	l \sum_{\alpha=1}^{N}x_{\alpha,i}\frac{\partial V_{l}}{\partial x_{\alpha,i}}+V_{l}=0,\quad
	\sum_{\alpha=1}^{N}\left(x_{\alpha,i}\frac{\partial V_{l}}{\partial x_{\alpha,j}}-x_{\alpha,j}\frac{\partial V_{l}}{\partial x_{\alpha,i}}\right)=0.
	\eea
	
	It is easy to see that there is a correspondence between potentials $V_{\frac{1}{2}}$ and solutions of the system (\ref{sys2}).
	Indeed, let us suppose that we have a solution $V_{\frac{1}{2}}$ of the system (\ref{sys1}). Then we can produce a solution
	\bea
	V_{l}=\sqrt[2l]{V_{\frac{1}{2}}}
	\nonumber
	\eea
	of the system (\ref{sys2}) for any possible value of $l$. It should be noted that the potential  $V_{\frac{1}{2}}$ should be positive-definite in order to obtain a solution $V_{l}$ for integer $l$. For example, the potential
	\bea
	V_{l}=g^2\sqrt[2l]{\sum_{\alpha<\beta}\frac{1}{(x_{\alpha,i}-x_{\beta,i})^2}}
	\nonumber
	\eea
	is related to the celebrated Calogero model \cite{Calogero}.
	
	To obtain NH counterpart of the model (\ref{MB}), it follows to apply Niederer's coordinate transformation of the form
	\bea\label{Nied_1}
	t=R \tan{(\tau/R)},\quad x_{\alpha,i}(t)=\frac{\chi_{\alpha,i}(\tau)}{\cos^{2l}{(\tau/R)}},
	\eea
	to the action functional (\ref{MB}). For half-integer $l$ one finds
	\bea\label{PUev1}
	S=\frac{1}{2}\int d\tau\left[ \,\sum_{\alpha=1}^{N}\chi_{\alpha,i}\prod_{k=0}^{l-1/2}\left(\frac{d^2}{d\tau^2}+\frac{(2k+1)^2}{R^2}\right)\chi_{\alpha,i}-
	V_{l}(\chi_{1,i},\chi_{2,i},..,\chi_{N,i})\right] ,
	\eea
	while for integer $l$ one obtains
	\bea\label{PUod1}
	S=\frac{1}{2}\int d\tau\left[ \,\sum_{\alpha=1}^{N}\epsilon_{ij}\,\chi_{\alpha,i}\prod_{k=1}^{l}\left(\frac{d^2}{d\tau^2}+\frac{(2k)^2}{R^2}\right)\dot{\chi}_{\alpha,j}-
	V_{l}(\chi_{1,i},\chi_{2,i},..,\chi_{N,i})\right] ,
	\eea
	where the function $V_{l}=V_{l}(\chi_{1,i},\chi_{2,i},..,\chi_{N,i})$ obeys the same conditions as in (\ref{sys2})
	\bea
	\sum_{\alpha=1}^{N}\frac{\partial V_{l}}{\partial \chi_{\alpha,i}}=0,\quad
	l \sum_{\alpha=1}^{N}\chi_{\alpha,i}\frac{\partial V_{l}}{\partial \chi_{\alpha,i}}+V_{l}=0,\quad
	\sum_{\alpha=1}^{N}\left(\chi_{\alpha,i}\frac{\partial V_{l}}{\partial \chi_{\alpha,j}}-\chi_{\alpha,j}\frac{\partial V_{l}}{\partial \chi_{\alpha,i}}\right)=0.
	\nonumber
	\eea
	The actions (\ref{PUev1}) and (\ref{PUod1}) describe a set of identical Pais-Uhlenbeck oscillators which interact with each other via a conformal-invariant potential. This system 
	can be viewed as a higher-derivative generalization of many-body conformal mechanics considered in \cite{Galajinsky_8,Galajinsky_9}.
	
	\vskip 1cm
	\noindent
	{\bf 5. Conclusion}
	\vskip 0.5cm
	To summarize, in this work higher-derivative analogues of the conformal particle and a system of
	identical particles which interact with each other via a conformal-invariant potential were constructed.  An appropriate Niederer's coordinate transformation was applied
	to these higher-derivative systems so as to obtain their NH counterparts.
	
	Turning to further possible developments, the most interesting questions are related to integrability and stability. A construction of
	supersymmetric extensions of higher-derivative models (\ref{CHP}), (\ref{CPU}), (\ref{MB}), (\ref{PUev1}),
	and (\ref{PUod1}) is worth studying as well. It would be also interesting to extend the analysis in
	Refs. \cite{Galajinsky_9,Galajinsky_5,Galajinsky_6,Galajinsky_7}, which is related to nonlocal conformal transformations,
	to the case of higher-derivative mechanical systems introduced in this paper.

\begin{acknowledgments}
	The author would like to express his gratitude to I. Masterov for posing the problem and useful discussions and to A. Galajinsky for his useful comments.
	This work was supported by the RF Presidential grant MK-2101.2017.2.
\end{acknowledgments}	
	\fontsize{10}{13}\selectfont


\begin{thebibliography}{nn}
		\bibitem{Niederer_1}
		U. Niederer, \\ {\it The maximal kinematical invariance group of the free Schrodinger equation}, \\ Helv. Phys. Acta {\bf 45} (1972) 802.
		\bibitem{Alfaro}
		V. de Alfaro, S. Fubini, G. Furlan,\\ {\it Conformal invariance in quantum mechanics},\\ Nuovo Cim. A {\bf 34} (1976) 569-612.
		\bibitem{Niederer}
		U. Niederer,\\ {\it The maximal kinematical invariance group of the harmonic oscillator},\\ Helv. Phys. Acta {\bf 46} (1973), 191-200.
		\bibitem{Calogero}
		F. Calogero,\\ {\it Solution of the one-dimensional N body problems with quadratic and/or inversely quadratic pair potentials}, J. Math. Phys. {\bf 12} (1971) 419-436.
		
		
		
		\bibitem{Johnson}
		N.F. Johnson, L. Quiroga,\\ {\it Analytic results for N particles with 1/r2 interaction in two dimensions
			and an external magnetic field},\\ Phys. Rev. Lett. {\bf 74} (1995) 4277-4280.
		\bibitem{Gibbons}
		G.W. Gibbons, P.K. Townsend,\\ {\it Black holes and Calogero models},\\ Phys. Lett. B {\bf 454} (1999) 187-192, hep-th/9812034.
		
		\bibitem{Horvathy}
		C. Duval, P. Horv\'{a}thy,\\ {\it Non-relativistic conformal symmetries and Newton-Cartan structures},\\ J. Phys. A {\bf 42} (2009) 465206, arXiv:0904.0531 [math-ph].
		
		\bibitem{Henkel}
		M. Henkel, \\ {\it Local scale invariance and strongly anisotropic equilibrium critical system}, \\ Phys. Rev. Lett. {\bf 78} (1997) 1940, cond-mat/9610174.
		\bibitem{Negro_1}
		J. Negro, M.A. del Olmo, A. Rodriguez-Marco, \\ {\it Nonrelativistic conformal groups}, \\ J. Math. Phys. {\bf 38} (1997), 3786.
		\bibitem{Negro_2}
		J. Negro, M.A. del Olmo, A. Rodriguez-Marco, \\ {\it Nonrelativistic conformal groups. II. Further developments and physical applications}, \\ J. Math. Phys. {\bf 38} (1997), 3810.
		\bibitem{Lukierski}
		J. Lukierski, P.C. Stichel, W.J. Zakrzewski, \\ {\it Exotic Galilean conformal symmetry and its dynamical realisations}, \\ Phys. Lett. A {\bf 357} (2006), 1-5.
		\bibitem{Galajinsky_3}
		A. Galajinsky, I. Masterov, \\ {\it Remark on l-conformal extension of the Newton-Hooke algebra}, \\ Phys.Lett. B {\bf 702} (2011), 265-267, arXiv:1104.5115 [hep-th].
		
		\bibitem{DH}
		C. Duval, P.A. Horvathy,\\ {\it Conformal Galilei groups, Veronese curves, and Newton-Hooke spacetimes},\\ J. Phys. A {\bf 44} (2011), 335203, [arXiv:1104.1502].
		\bibitem{Gomis}
		J. Gomis, K. Kamimura, \\ {\it Schrodinger equations for higher order non-relativistic particles and N-Galilean conformal symmetry}, \\ Phys. Rev. D {\bf 85} (2012), 045023, arXiv:1109.3773 [hep-th].
		\bibitem{Pais}
		A. Pais, G.E. Uhlenbeck,\\ {\it On field theories with nonlocalized action},\\ Phys. Rev. {\bf 79} (1950) 145-165.
		\bibitem{Galajinsky_1}
		A. Galajinsky, I. Masterov,\\ {\it Dynamical realizations of l-conformal Newton-Hooke group},\\ Phys. Lett. B {\bf 723} (2013) 190, arXiv:1303.3419.
		\bibitem{AG}
		K. Andrzejewski, J. Gonera,\\ {\it Dynamical interpretation of nonrelativistic conformal groups},\\ Phys. Lett. B {\bf 721} (2013), 319-322.
		\bibitem{PU}
		K. Andrzejewski, A. Galajinsky, J. Gonera, I. Masterov,\\ {\it Conformal Newton-Hooke symmetry of Pais-Uhlenbeck oscillator},\\ Nucl. Phys. B {\bf 885} (2014) 150-162, arXiv:1402.1297 [hep-th].
		\bibitem{Andrzejewski}
		K. Andrzejewski,\\ {\it Conformal Newton-Hooke algebras, Niederer's transformation and Pais-Uhlenbeck oscillator},\\ Phys. Lett. B {\bf 738} (2014) 405-411, arXiv:1409.3926 [hep-th].
		\bibitem{Andrzejewski_1}
		K. Andrzejewski,\\ {\it Hamiltonian formalisms and symmetries of the Pais-Uhlenbeck oscillator},\\ Nucl. Phys. B {\bf 889} (2014) 333-350, arXiv:1410.0479 [hep-th].
		\bibitem{Masterov_2}
		I. Masterov,\\ {\it Dynamical realizations of N=1 l-conformal Galilei superalgebra},\\ J. Math. Phys. {\bf 55} (2014) 102901, arXiv:1407.1438 [hep-th].
		\bibitem{AB}
		A. Galajinsky, I. Masterov,\\ {\it On dynamical realizations of l-conformal Galilei and Newton-Hooke algebras},\\ Nucl. Phys. B {\bf 896} (2015) 244-254, arXiv:1503.08633 [hep-th].
		\bibitem{Andr}
		K. Andrzejewski,\\ {\it Generalized Niederer's transformation for quantum Pais-Uhlenbeck oscillator},\\ Nucl. Phys. B {\bf 901} (2015) 216-228, arXiv:1506.04909 [hep-th].
		
		\bibitem{Coleman_1}
		S.R. Coleman, J. Wess, B. Zumino,\\ {\it Structure of phenomenological Lagrangians. 1.},\\ Phys. Rev. {\bf 177} (1969) 2239.
		\bibitem{Coleman_2}
		C.G. Callan, S.R. Coleman, J. Wess, B. Zumino,\\ {\it Structure of phenomenological Lagrangians. 2.},\\ Phys Rev. {\bf 177} (1969) 2247.
		\bibitem{Ivanov}
		E.A. Ivanov, V.I. Ogievetsky, \\ {\it The inverse Higgs phenomenon in nonlinear realizations},\\ Teor. Mat. Fiz. {\bf 25} (1975) 164.
		\bibitem{Ivanov_1}
		E.A. Ivanov, S.O. Krivonos, V.M. Leviant,\\ {\it Geometry of conformal mechanics},\\ J. Phys. A {\bf 22} (1989) 345.
		
		\bibitem{Galajinsky_8}
		A. Galajinsky,\\ {\it N=2 superconformal Newton-Hooke algebra and many-body mechanics},\\ Phys. Lett. B {\bf 680} (2009) P. 510-515, arXiv:0906.5509[hep-th].
		\bibitem{Galajinsky_9}
		A. Galajinsky,\\ {\it Conformal mechanics in Newton-Hooke spacetime},\\ Nucl. Phys. B {\bf 832} (2010) P. 586-604, arXiv:1002.2290[hep-th]. 
		
		\bibitem{Galajinsky_5}
		A. Galajinsky, O. Lechtenfeld, K. Polovnikov,\\ {\it Calogero models and nonlocal conformal transformations},\\ Phys. Lett. B {\bf 643} (2006) P. 221-227, hep-th/0607215.
		\bibitem{Galajinsky_6}
		A. Galajinsky,\\ {\it Remark on quantum mechanics with conformal Galilean symmetry},\\ Phys. Rev. D {\bf 78} (2008) 087701, arXiv:0808.1553 [hep-th].
		\bibitem{Galajinsky_7}
		A. Galajinsky, I. Masterov,\\ {\it Remark on quantum mechanics with N=2 Schrodinger supersymmetry},\\ Phys. Lett. B {\bf 675} (2009) P. 116-122, arXiv:0902.2910 [hep-th].
		
		
		
		
		
	\end{thebibliography}
\end{document}